\documentclass[article,singlecolumn,superscriptaddress,preprintnumbers,nopacs,floatfix,amsmath,amssymb]{revtex4}

\usepackage{graphicx}
\usepackage{dcolumn}
\usepackage{bm}

\usepackage{amsmath}
\usepackage{amssymb}
\usepackage{mathrsfs}
\usepackage{bbm}
\usepackage{epsfig}
\usepackage[mathscr]{euscript}
\usepackage{calligra}
\usepackage[T1]{fontenc}
\usepackage{aurical}
\usepackage[T1]{fontenc}
\usepackage{wrapfig}
\usepackage{eurosym}
\usepackage[usenames,dvipsnames]{color}
\usepackage[yyyymmdd,hhmmss]{datetime}

\def\nn{\nonumber}

\newcommand{\be}{\begin{eqnarray}}
\newcommand{\ee}{\end{eqnarray}}
\newcommand{\br}{\begin{matrix}}

\newcommand{\er}{\end{matrix}}

\newcommand{\Z}{{Z \!\!\! Z}}

\renewcommand{\vector}[1]{{\boldsymbol{#1}}}
\newcommand{\eastar}{\end{eqnarray*}}

\begin{document}

\title{Poisson Hierarchy of Discrete Strings}

\author{Theodora Ioannidou}
\email{
ti3@auth.gr
}
\affiliation{
Faculty of Civil Engineering,  School of Engineering,  Aristotle University of Thessaloniki, 54249, Thessaloniki, Greece
}
\author{Antti J. Niemi}
\email{Antti.Niemi@physics.uu.se}
\affiliation{Department of Physics and Astronomy, Uppsala University,
P.O. Box 803, S-75108, Uppsala, Sweden}
\affiliation{
Laboratoire de Mathematiques et Physique Theorique
CNRS UMR 6083, F\'ed\'eration Denis Poisson, Universit\'e de Tours,
Parc de Grandmont, F37200, Tours, France}
\affiliation{Department of Physics, Beijing Institute of Technology, Haidian District, Beijing 100081, P. R. China\\}

\begin{abstract}
\noindent
The Poisson geometry of a discrete string in three dimensional Euclidean space is 
investigated. For this the Frenet frames are  converted  into a  spinorial representation, 
the discrete
spinor Frenet equation is interpreted in terms of a  transfer matrix  formalism, and Poisson brackets are
introduced in terms  of the spinor components.  The construction is then
generalised, in a self-similar manner, into an infinite hierarchy of Poisson algebras. 
As an example, the classical Virasoro (Witt) algebra that determines reparametrisation 
diffeomorphism along a continuous string,  is identified as a particular sub-algebra, 
in the hierarchy of the discrete string Poisson algebra.
\end{abstract}

\maketitle

A continuous curve in the three dimensional Euclidean space $\mathbb R^3$  
is a classic subject of differential geometry \cite{spivak}. 
The study of curves in $\mathbb R^3$ is similarly pivotal in physics  
where plenty of breakthroughs come with strings attached. This  
includes in particular topics that relate to the theory knots  \cite{Kauffman-1991} such as
topological Chern-Simons theories \cite{Witten-1989}, 
knotted solitons \cite{Faddeev-1997} and exotic exchange statistics \cite{wilczek}. 
In the present Letter, the Poisson geometry of a {\it discrete} string in ambient $\mathbb R^3$ is 
studied. In particular, a novel
infinite dimensional, self-similar hierarchy of Poisson bracket algebras  is exposed. 
Such a hierarchy is important, for example in the construction of  integrable 
Hamiltonian models of discrete string dynamics \cite{Hu-2013,Ioannidou-2014}. It also 
facilitates the numerical study of continuous strings {\it e.g.} on a discrete lattice.
Moreover, the concept of a discrete string is a most useful one
to aspects of computer graphics, virtual reality and robotics \cite{hansonbook,kuipers}. 
Discrete strings also model polymers \cite{schafer}, including biophysical ones from DNA 
to proteins \cite{Danielsson-2010,Chernodub-2010,Molkenthin-2011,Niemi-2014}. 

Our starting point is the description of a discrete string in terms of an open and oriented, 
piecewise linear polygonal chain
${\bf r}(s) \in \mathbb R^3$ \cite{Hu-2011}. The arc length  parameter  takes values on 
$s \in [0,L]$ where $L$ is the total length of the string. 
The  vertices ${\mathscr D}_i$ that specify the string  are located at the points
${\bf r}_i = ({\bf r}_0, \dots , {\bf r}_n)$  with ${\bf r}(s_i) = {\bf r}_i$;
the  endpoints of the string are  ${\bf r}(0) = {\bf r}_0$ and  ${\bf r}(L) = {\bf r}_n$.
In addition, the distance of    the nearest neighbour vertices ${\mathscr D}_{i}$ and ${\mathscr D}_{i+1}$  is 
\[
|\mathbf r_{i+1} - \mathbf r_i | = s_{i+1} - s_i.
\]
Therefore, the  nearest neighbour vertices are  connected by the line segments
\[
{\bf r}(s) \ = \ \frac{ s-s_{i} } {s_{i+1} - s_i} \, {\bf r}_{i+1}  \ - \ \frac{ s - s_{i+1}}  {s_{i+1} - s_i} \, {\bf r}_{i} ,
\ \ \ \ \ \ (s_{i} < s < s_{i+1}).
\]
The unit length discrete tangent vector $\mathbf t_i = $($t_1^i, t_2^i, t_3^i$) 
that points from vertex ${\mathscr D}_{i}$ to vertex ${\mathscr D}_{i+1}$ is defined as
\begin{equation}
\mathbf t_i \ = \ \frac{ \mathbf r_{i+1} - \mathbf r_i }{ |\mathbf r_{i+1} - \mathbf r_i |}.
\label{defti}
\end{equation}
Thus, the vertex ${\mathscr D}_{k}$ is located at a point
\begin{equation}
\mathbf r_k \ = \ \sum\limits_{i=0}^{k-1} \ \left| \mathbf r_{i+1} - \mathbf r_i \right| \, \mathbf t_i.
\label{rtstring}
\end{equation}

We are interested in the Poisson geometry and the ensuing algebraic structures
that can be associated to such a three dimensional discrete  string. 
In lieu of the traditional approach  which is 
based on the discrete Frenet frames 
in terms of the tangent, normal and binormal vectors,  we utilise  
a two component complex spinor description, with the spinors supported along the string \cite{Ioannidou-2014}. 
Then the discrete Frenet equation becomes a two component spinor 
Frenet equation  \cite{Ioannidou-2014}.  Such a spinor based representation of the discrete 
string geometry has already been found to have various conceptual and technical advantages,
including the relations between the time evolution of the discrete string and known integrable equations. For 
more details, see Ref. \cite{Ioannidou-2014}. 

We proceed as follows: To each link from vertex ${\mathscr D}_i$ to vertex  ${\mathscr D}_{i+1}$
we associate a two component complex spinor 
\begin{equation}
\psi_{i} =  \left( \begin{matrix} z^{i}_{1}  \\ z^{i}_{2} \end{matrix}\right),
\label{psi1}
\end{equation}
where the $z^{i}_{a}$ (for $a=1,2$ and $i\in\Z$) are complex variables with support 
on the link. The spinors are related to the unit length tangent vectors 
by
\begin{equation}
\sqrt{g_i}\, {\mathbf t}_i  =
<\psi_i, \hat \sigma \psi_i> .
\label{tpsii}
\end{equation}
The  $\hat \sigma = (\sigma^1, \sigma^2, \sigma^3)$ are the standard Pauli matrices and
\begin{equation}
\sqrt{g_i} \equiv  |z^{i}_{1}|^2 + |z^{i}_{2 }|^2
\label{gscale}
\end{equation}
is a metric scale factor. 
Explicitly,
\begin{equation}
\sqrt{g_i} \left( \begin{matrix} t^i_1+ i t^i_2  \\ t^i_3 \end{matrix}\right) \ = \ 
\left( \begin{matrix} 2 \bar z^{i}_{1}  z^{i}_{2} \\  
\bar z^{i}_{1}  z^{i}_{1}   
- \bar z^{i}_{2} z^{i}_{2} \end{matrix} \right).
\label{tz1z2}
\end{equation}
Together with (\ref{defti}) this
determines the spinor components ($z_1^i, z_2^i$) in terms of the vertices $\mathcal D_i$, up to an overall phase.
In addition, for each $i$ the conjugate spinor $\bar\psi_i$ is  defined by introducing 
the charge conjugation operation ${\mathscr C}\,$ that acts on $\psi_i$
in the following way
\begin{equation}
{\mathscr C} \,  \psi_i \ = \  -i \sigma_2 \psi_i^\star = \bar \psi_i \ = \ \left( \begin{matrix} - \bar z^{i}_{2} 
\\ \  \ \bar z^{i}_{1} \end{matrix}\right).
\label{C}
\end{equation}
Observe that 
\[
\mathscr C^2 = - \mathbb I
\]
and  that the two spinors are orthogonal since
\[
<\psi_i \, , \bar\psi_i> = 0.
\]

We combine the spinor components  into a $2\times 2$ matrix  $\mathfrak u_i$
as follows:
\begin{equation}
{\mathfrak u}_i \  = \  \left( \begin{matrix} z^{i}_{1} & -\bar z^{i}_{2} \\ z^{i}_{2} & \ \ \bar 
z^{i}_{1} \end{matrix} \right),
\label{g}
\end{equation}
so that
\begin{eqnarray*}
\psi _i  =   {\mathfrak u}_i  
\left( \begin{matrix} 1 \\ 0 \end{matrix} \right) \  \ \ \ \ \ \ \& \ \ \ \ \ \ \ 
 \bar \psi_i  =   {\mathfrak u}_i  
\left( \begin{matrix} 0 \\ 1 \end{matrix} \right).
\end{eqnarray*}
Our main observation  is, that in the case of  an infinite number of vertices ${\mathscr D}_i$
this leads to an infinite hierarchy of spinors and ensuing discrete strings and 
Poisson algebras. Iteratively, and in a self-similar manner; 
for a finite number of vertices we obtain a finite dimensional
sub-hierarchy.  

To expose the hierarchy together with its self-similar structure, we start by defining a four-component spinor obtained by combining  the two spinors into a Majorana spinor
\begin{equation}
\Psi_i = \left( \begin{matrix} - \bar \psi_i \\ \ \ \psi_i \end{matrix} \right).
\label{majspi}
\end{equation}
Indeed,  under conjugation by $\mathscr C$ this four component spinor  transforms according to 
\[
\Psi_i \ \longrightarrow \ \mathscr C \, \Psi_i   \ = \ - i \sigma_2 \Psi^\star_i  \ = \  \left( \begin{matrix} \psi_i \\ \bar \psi_i \end{matrix} \right)  \equiv \ 
\left( \begin{matrix} \ \ 0 & 1  \\ - 1 & 0 \end{matrix} \right) \left( \begin{matrix} - \bar \psi_i \\  \ \ \psi_i \end{matrix} \right)
\]
where  the Pauli matrices  $\sigma_a$ (now) act  in the two dimensional
space of  the spinor components of $\Psi_i$. In terms of these Majorana spinors, 
the original discrete spinorial Frenet equation takes the following form
\begin{equation}
\Psi_{i+1} \ = \  \mathcal U_{i} \Psi_i.
\label{discfre1}
\end{equation}
Here $\mathcal U_{i} $ is the ensuing transfer matrix, in the chain of spinors $\Psi_i$.
The self-similar structure emerges when we parametrise $\mathcal U_{i} $
{\it exactly}  in accordance with (\ref{g}). 
That is, by setting
\begin{equation}
\mathcal U_{i} \ = \ \left( \begin{matrix}  Z^{i} _{1} 
& - \bar Z^{i}_{2}\\   Z^{i}_{2}& \ \ \bar Z^{i}_{1}\end{matrix} \right).
\label{matU}
\end{equation}
Then a relation between the variables $(z^i_1,z^i_2)$ and $(Z^i_1,Z^i_2)$ is  obtained from (\ref{discfre1}) since
\begin{equation}
\sqrt{g_i}\ \mathcal U_{i}=\Psi_{i+1}\Psi_i^\dagger.
\label{deftU}
\end{equation}
This implies that
\begin{eqnarray}
\sqrt{g_i} Z_{1}^{i} &=& \bar z_{1}^{i+1} z_{1}^{i} + \bar z_{2}^{i+1} z_{2}^{i}  \nn \\ 
\sqrt{g_i} Z_{2}^{i} &=&  z_{1}^{i+1} z_{2}^{i}  - z_{2}^{i+1} z_{1}^{i}.
\label{Z}
\end{eqnarray}
%
%
%
%
%
%
%
For each pair of indices  ($k,l$) we introduce the
variables $W_1^{k,l}$ and $W_2^{k,l}$ as 
\begin{eqnarray}
W_1^{k,l}& \buildrel {def}\over{=} & \bar z_{1}^{k}\, z_{1}^{l} + \bar z_{2}^{k} \, z_{2}^{l}  \nn \\ 
W_2^{k,l}& \buildrel {def}\over{=} &  z_{1}^{k} \, z_{2}^{l}  - z_{2}^{k} \, z_{1}^{l},
\label{W}
\end{eqnarray}
so that in particular,
\begin{eqnarray*}
W_1^{i+1,i} & \equiv \sqrt{g_i} \,Z_{1}^{i} \\
W_2^{i+1,i} & \equiv\sqrt{g_i}\, Z_{2}^{i}.
\end{eqnarray*}

Note that the variables ($Z_{1}^i$, $Z_2^i$)  are the initiator variables of our self-similar hierarchy, 
{\it i.e.} they are the first level variables of the hierarchy. The 
variables $z_{1}^i$ and $z_2^i$ comprise the second level variables. The next level of hierarchy
then emerges when, in analogy with (\ref{Z}) and (\ref{W}), we proceed by setting
\begin{eqnarray}
\sqrt{ {\mathfrak g}_i }  z_{1}^{i} &=& \bar {\mathfrak z}_{1}^{i+1}\,  {\mathfrak z}_{1}^{i} + \bar {\mathfrak z}_{2}^{i+1} 
{\mathfrak z}_{2}^{i} \ \buildrel {def}\over{=} w_1^{i+1,i} \nn \\
\sqrt{ {\mathfrak g}_i }  z_{2}^{i} &=&  {\mathfrak z}_{1}^{i+1} \, {\mathfrak z}_{2}^{i} -  {\mathfrak z}_{2}^{i+1} 
{\mathfrak z}_{1}^{i} \ \buildrel {def}\over{=} w_2^{i+1,i}
\label{frakz}
\end{eqnarray}
where the metric scale is
\[
\sqrt{ {\mathfrak g}_i } \ = \ |\mathfrak z_1^{i}|^2 + |{\mathfrak z}^{i}_{2}|^2.
\]
In analogy with (\ref{psi1}), we then introduce the two component spinors 
\[ 
\chi_i \ = \ \left( \begin{matrix} {\mathfrak z}_{1}^{i} \\ {\mathfrak z}_{2}^{i} \end{matrix} \right)
 \ \ \ \ \  \& \ \ \ \ \ 
\bar{ \chi}_i \ = \ \left( \begin{matrix} -\bar{{\mathfrak z}}_{2}^{i} \\ \  \ \bar{{\mathfrak z}}_{1}^{i} \end{matrix} \right)
\]
which are combined into the following four component Majorana
spinors  (in accordance with (\ref{majspi})):
\begin{equation}
\mathcal X_i  \ = \ \left( \begin{matrix} - \bar \chi_i \\ \ \  \chi_i \end{matrix} \right).
\label{Xi}
\end{equation}
%
%
In a self-similar repeat of the previous construction,  
the Majorana spinors (\ref{Xi}) are then related to each other by an equation that has the transfer matrix form 
\begin{equation}
\mathcal X_{i+1} \ = \  \mathfrak u_{i} \mathcal X_i
\label{littleu}
\end{equation}
where $\mathfrak u_i$ is defined in (\ref{g}).
In analogy with (\ref{deftU}), this is the second level transfer matrix which implies that
\[
\sqrt{ {\mathfrak g}_i } \, \mathfrak u_{i} \ = \ \mathcal X_{i+1} \mathcal X_{i}^\dagger.
\]
The aforementioned construction can be extended to higher levels, in a straightforward self-similar manner.  
In this way we obtain the following infinite self-similar hierarchy of variables   
\begin{equation}
Z_{a}^{i} \ \buildrel{ \mathcal U_i} \over \longrightarrow \ z_{a}^{i} \ \buildrel{ \mathfrak u_i} \over \longrightarrow   
{\mathfrak z}_{a}^{i}
 \ \buildrel{ \mathfrak v_i} \over \longrightarrow \ \dots \ \ \ \ \ \ (a=1,2),
\label{setvar}
\end{equation}
mapped onto each other by the ensuing transfer matrices $\mathcal U_i , \, \mathfrak u_i , \mathfrak v_i, \dots$ as in 
(\ref{discfre1}), (\ref{littleu})  and so forth; note that the transfer matrices always have the same functional form 
(\ref{g}), (\ref{matU}) in the respective variables.

Therefore,   for each set of variables in the hierarchy, we can introduce 
 the corresponding  piecewise linear discrete string. For this we use the ensuing relation (\ref{tz1z2}) 
between the variables and the tangent vector of the corresponding string (\ref{rtstring}).
For example, in the case of  the $Z_{a}^{i}$  (similarly to 
(\ref{tpsii})) we have
\[
\sqrt{G_i} \left( \begin{matrix} T^i_1+ i T^i_2  \\ T^i_3 \end{matrix}\right) \ = \  
\left( \begin{matrix} 2 \bar Z^{i}_{1} Z^{i}_{2} \\ \bar Z^{i}_{1} Z^{i}_{1}   
- \bar Z^{i}_{2} Z^{i}_{2} \end{matrix} \right)
\]
with the metric scale given by
\[
\sqrt{G_i} = |Z_1^i|^2 + |Z_2^i|^2
\]
while the vertices of the corresponding string are located at the points
\begin{equation}
\mathbf R_k \ = \ \sum\limits_{i=0}^{k-1}\ | \mathbf R_{i+1} - \mathbf R_i | \cdot \mathbf T_i.
\label{RTstring}
\end{equation}
We also note that the entire hierarchy of strings can  be framed, in a self-similar manner, using the following procedure
at each level of hierarchy:
We recall (\ref{g}) to introduce the matrices
\begin{eqnarray}
\hspace{-10mm} {\mathfrak u}_i^{-1}  \sigma_3 \, {\mathfrak u}_i 
&=&{ \mathbf t^i} \cdot {\vector\sigma} \ \equiv \ \hat {{\mathbf t}}^i
\label{gmat1}
\\
\hspace{-15mm}{\mathfrak u}_i^{-1} \sigma_\pm \, {\mathfrak u}_i &\equiv 
& \frac{1}{2}{\mathfrak u}_i^{-1} \! \left(\sigma_1 \pm i \sigma_2 \right) {\mathfrak u}_i
\ = \mathbf e^i_\pm \cdot {\vector \sigma} \ \equiv \ \hat{\mathbf  e}^i_\pm,
\label{gmat2}
\end{eqnarray}
in terms of the transfer matrix ${\mathfrak u}_i$.  These matrices obey the \underline{su}(2) Lie algebra
\begin{equation}
[ \, \hat{\mathbf  t^i} \, , \, \hat{\mathbf e}^i_\pm \, ] \ = \  \pm 2 \hat{\mathbf  e}^i_\pm, 
\ \ \ \ \ \ \ \ \ \ \ \ 
[ \, \hat{\mathbf  e}^i_+ \, , \, \hat{\mathbf  e}^i_- \, ] \ = \   \hat{\mathbf t}^i.
\label{su2}
\end{equation}
The  ($\hat{\mathbf  t}^i, \hat{\mathbf  e}^i_\pm $) then define 
a generic right-handed orthonormal frame at vertex ${\mathscr D}_i$.
A frame rotation that leaves $\hat{\mathbf  t}^i$ intact, acts by  a 
${\calligra h_i}\in$ U(1) $\!\!\subset\!$ SU(2) multiplication  
of ${\mathfrak u}_i$ from the left.  That is,  by letting 
\begin{equation}
{\mathfrak u}_i \ \buildrel {\calligra h}_i \over \longrightarrow \ {\calligra h}_i {\mathfrak u}_i,
 \hspace{5mm}  \ \ 
{\calligra h}_i \, = \, e^{i\varphi_i \sigma_3}.
\label{gh}
\end{equation}
we have
\begin{eqnarray}
\hat{\mathbf  t}^i &\buildrel {\calligra h_i} \over \longrightarrow & {\mathfrak u}_i ^{-1} {\calligra h}_i^{-1} \sigma_3 
{\calligra h}_i {\mathfrak u}_i
\equiv \hat{\mathbf  t}^i
\label{3nro}\\
\hat{\mathbf  e}^i_{\pm} & \buildrel {\calligra h}_i \over \longrightarrow &
{\mathfrak u}_i^{-1}{\calligra h}_i^{-1}\sigma^\pm {\calligra h}_i{\mathfrak u}_i \ = \ e^{\pm 2i\varphi_i} \hat{\mathbf e}^i_\pm.
\label{3ero}
\end{eqnarray}
Analogous relations can be introduced, for all levels of the hierarchy, in terms of the ensuing transfer matrices. 

We now proceed to reveal our infinite self-similar hierarchy of Poisson algebras, defined in terms of the  
symplectic structures of the variables in the hierarchy. To start the 
construction, we  impose  the following Poisson bracket at the second  level of the hierarchy:
\begin{equation}
\{ z_{a}^{j} , \bar z_{b}^{k} \} \ = \ i \, \omega(k)\,\delta_{ab}^{jk}.
\label{bracket1}
\end{equation}
We assume that all the remaining brackets between the ($z_a^i ,  z_b^j$) variables and their 
conjugates vanish; this clearly defines a symplectic structure, for the local coordinates ($z_a^i ,  z_b^j$).
Note that,  for the  canonical Heisenberg algebra  $\omega(k) \equiv 1$ and 
 for the components (\ref{tz1z2})  of the tangent vectors,  the Poisson brackets are given by
\begin{equation}
\{ \sqrt{g_i} \, t_a^i , \sqrt{g_j} \, t_b^j \} \, =  \, \epsilon_{abc} \delta^{ij} \sqrt{g_i} t_c^i .
\label{ttt}
\end{equation} 
The self-similar structure then gives us an infinite hierarchy of Poisson algebras, as follows: We simply
substitute the ensuing variables in the hierarchy, into the Poisson bracket relation such as (\ref{bracket1})
which is expressed in terms of the preceding variables in the hierarchy.
This yields the brackets between all the variables at all  levels of  hierarchy, order-by-order. 
Let us consider,  as an example, the algebra that  we obtain for the variables defined by the equations (\ref{W});
note that for this, we have to proceed in the opposite direction, from the second level down to the first.
It is straightforward to show that  the (only) non-vanishing brackets of two variables (\ref{W}), when located  at the
{\it same site}, are given by
\begin{eqnarray}
\{ W_{1}^{i+1,i} , \bar W_{1}^{i+1,i} \} &=& i \left( \omega(i)\,  \sqrt{g_{i+1}} - \omega(i+1) \,\sqrt{g_{i}} \right)
\label{W1W1}\\
\{ W_{2}^{i+1,i} , \bar W_{2}^{i+1,i} \} &=& i \left( \omega(i)\, \sqrt{g_{i+1}} +  \omega(i+1) \,\sqrt{g_{i}} \right).
\label{W1W2}
\end{eqnarray}
Note that  (\ref{W1W1}) vanishes when
\[
\frac{ \omega(i+1) }{\omega(i)} = \frac{  \sqrt{g_{i+1}} }{\sqrt{g_{i}}}.
\]
In addition,  (\ref{W1W2})  becomes the  Heisenberg bracket when
\[
\omega(i) \,\sqrt{g_{i+1}} +  \omega(i+1) \,\sqrt{g_{i}}  =1.
\]
Furthermore, for a pair of variables (\ref{W}) which are located at {\it different sites} (not 
necessarily nearest   neighbour)   the only  non-vanishing brackets are the following ones
\begin{eqnarray*}
\left\{W_2^{k,l},\bar{W}_2^{m,n}\right\}&=&i\omega\left(k\right)\,\delta^{km}\,W_1^{n,l}+i\omega\left(l\right)\,\delta^{ln}\,W_1^{m,k}
-i\omega(l)\,\delta^{lm}\,W_1^{n,k}-i\omega(k)\,\delta^{kn}\,W_1^{m,l}\nonumber\\
\left\{W_1^{k,l},W_1^{m,n}\right\}&=&i\omega\left(l\right)\,\delta^{lm}\,W_1^{k,n}-i\omega\left(k\right)\,\delta^{kn}\,W_1^{m,l}\nonumber\\
\left\{W_2^{k,l},W_1^{m,n}\right\}&=&i\omega\left(k\right)\,\delta^{km}\,W_2^{n,l}+i\omega\left(l\right)\,\delta^{lm}W_2^{k,n},
\end{eqnarray*}
together with their complex conjugated brackets. This is the Poisson algebra, at the first level of the hierarchy.
Observe that by definition, due to (\ref{W}),  the $W_i^{k,l}$ (for $i=1,2$) 
satisfy the following identities
\begin{eqnarray}
 W_1^{k,l}= \bar{W}_1^{l,k},
\ \ \ \ \ \ \ \ \ \ W_2^{k,l}=-W_2^{l,k}.\label{proper}
\end{eqnarray}
Let us finally show how to identify the classical Virasoro generators in terms of these variables.
For this we assume  that the chain is  {\it infinitely} long.  We introduce the combinations
\begin{equation}
L^1_n = i \!\! \sum\limits_{k=-\infty}^\infty \!\! \bar z_1^{k} \,z_1^{k-n} \ \ \ \ \ \  \& \ \ \ \ \ \ L^2_n 
= i \!\! \sum\limits_{k=-\infty}^\infty \!\! \bar z_2^{k} \,z_2^{k-n},
\label{viragens}
\end{equation}
where $k\in\mathbb Z$.
Then the Poisson brackets of the $L_n^a$ are given by
\[
\{ L^a_n , L^b_m \} \ = \  i \!\! \sum\limits_{k=-\infty}^\infty \!\!  \left[ \omega(k-m) - \omega(k-n) \right]
\bar z_a^k z_a^{k-m-n}\delta^{ab}.
\]
By setting $\omega(k) = k$ we find that the variables $L_n^a$ satisfy the classical Virasoro (Witt) algebra.
That is,
\[
\{ L^a_n , L^b_m \} \ = \  i \left(n-m\right) \delta^{ab}  L^a_{n+m}.
\]
We also note that 
\[
L_n^1 + L_n^2 \ = \ i \!\! \sum\limits_{k=-\infty}^\infty \!\!  W_1^{k,k-n}. 
\]
Thus, in the infinite hierarchy that we have constructed in terms of  {\it discrete} strings, we have found 
the algebra of reparametrisations of {\it continuous} strings, as a sub-algebra.
Moreover, we have the following sub-algebra structure
\begin{eqnarray*}
\left\{L^1_n+L^2_n,W_1^{k,l}\right\}&=&-\omega\left(k\right)\,W_1^{k+n,l}+\omega\left(l\right)\,W_1^{k,l-n} \\
\left\{L^1_n+L^2_n,W_2^{k,l}\right\}&=&\omega\left(k\right)\,W_2^{k-n,l}+\omega\left(l\right)\,W_2^{k,l-n}\label{1}.
\end{eqnarray*}

\vskip 0.3cm
In summary, we have employed the spinorial formulation of a discrete string in combination with the formalism of discrete Frenet equations to derive an infinite algebraic Poisson hierarchy.  
As an example, we have shown that the structure of classical Virasoro (Witt) algebra of continuous strings becomes 
embedded in this hierarchy.  The structure we have revealed, forms a basis for studying the Poisson geometry of discrete strings which is the starting point for constructing integrable structures that 
model their dynamics in $\mathbb R^3$. 
\vskip 1.0cm

We both thank Y. Jiang and T.I., also,  thanks A. Doikou for discussions.  T.I. thanks 
IIP at Federal University of Rio Grande do Norte; 
Department of Physics and Astronomy at Uppsala University; and School  of Physics at 
Beijing Institute of Technology  for hospitality during the completion of this research. 
T.I. acknowledges  support from FP7, Marie Curie Actions, People, International Research Staff Exchange Scheme (IRSES-606096); and  from The Hellenic Ministry of Education: Education and Lifelong Learning Affairs, and European Social Fund: NSRF 2007-2013, Aristeia (Excellence) II (TS-3647).
A.J.N.  acknowledges  support from CNRS PEPS grant, Region Centre 
Rech\-erche d$^{\prime}$Initiative Academique grant; Sino-French Cai Yuanpei Exchange Program (Partenariat Hubert Curien), Vetenskapsr\aa det, Carl Trygger's Stiftelse f\"or vetenskaplig forskning; and  Qian Ren Grant at BIT.


\begin{thebibliography}{99}

\bibitem{spivak} M. Spivak, {\it A Comprehensive Introduction to Differential
Geometry} (Volume Two 3$^{rd}$ Ed.) (Publish or Perish, Inc, Houston, 1999 )
 
 \bibitem{Kauffman-1991} L. Kauffman, {\it Knots and Physics},  (World Scientific, Singapore, 1991) 
 
\bibitem{Witten-1989} E. Witten,  Commun. Math. Phys. {\bf 121} 351 (1989)
 
\bibitem{Faddeev-1997} L. Faddeev, A.J. Niemi, Nature {\bf 387} 58 (1997) 

\bibitem{wilczek} F. Wilczek, {\it Fractional Statistics and Anyon Superconductivity}, (World Scientific, Singapore, 1999)
 
\bibitem{Hu-2013} S. Hu, Y. Jiang, A.J. Niemi, Phys. Rev. D {\bf 87}  105011 (2013) 
 
\bibitem{Ioannidou-2014} T. Ioannidou, Y. Jiang, A.J. Niemi,  Phys Rev D {\bf 90} 025012  (2014)

\bibitem{hansonbook} A.J. Hanson, {\it Visualizing Quaternions}, Morgan Kaufmann Elsevier (London) 2006

\bibitem{kuipers} J.B. Kuipers, {\it Quaternions and Rotation Sequences: a Primer with Applications to Orbits, Aerospace, and Virtual Reality}, Princeton University Press (Princeton) 1999
 
 \bibitem{schafer} L. Sch\"afer, {\it Excluded Volume Effects in Polymer So-
lutions, as Explained by the Renormalization Group}
 (Springer Verlag, Berlin, 1999)

\bibitem{Danielsson-2010} U.H. Danielsson, M. Lundgren, A.J. Niemi, 
Phys.  Rev. E  {\bf 82}  021910 (2010) 

\bibitem{Chernodub-2010} M. Chernodub, S. Hu, A.J. Niemi, Phys. Rev. E {\bf 82} 011916 (2010) 

\bibitem{Molkenthin-2011}  N. Molkenthin, S. Hu, A.J. Niemi, Phys. Rev. Lett.  {\bf 106}  078102 (2011) 

\bibitem{Niemi-2014} A.J. Niemi, Theor. Math. Phys. {\bf 181} 1235 (2014)

\bibitem{Hu-2011} S. Hu, M. Lundgren, A.J. Niemi, Phys. Rev.  E {\bf 83}  061908 (2011)


\end{thebibliography}
\end{document}